\documentclass{mpe_report}

\usepackage{psfig,graphicx,epsfig}
\usepackage{color}
\usepackage{amsmath,amssymb,epic,eepic,array}

\begin{document}

\pagenumbering{arabic}
\setcounter{page}{124}

\renewcommand{\FirstPageOfPaper }{124}\renewcommand{\LastPageOfPaper }{127}%%
%% This is MPE-Report_example.tex
%% LaTeX2e example style file for the contributed talks and posters presented
%% during the 363rd Heraeus Seminar on Neutron Stars and Pulsars, held in 
%% Bad Honnef, May 14.-19. 2006.
%% 
%% This file needs the LaTeX2e class file he_symp.cls  
%%
% -----------------------------------------------------------------------------
%\documentclass{mpe_report}
%\usepackage{psfig}
% -----------------------------------------------------------------------------
%\def\R{~ROSAT}
%\def\RAS{\R all sky survey}
% -----------------------------------------------------------------------------
%\begin{document}

\title{Occurrence of concurrent `orthogonal' polarization modes in the Li\'enard-Wiechert field of 
 a rotating superluminal source}
\author{A.\ Schmidt\inst{1}, H.\ Ardavan\inst{2}, J.\ Fasel\inst{1}, J.\ Singleton\inst{1} \and A.\ Ardavan\inst{3}}  
\institute{Los Alamos National Laboratory, Los Alamos, NM 87545, USA
\and Institute of Astronomy, University of Cambridge, Cambridge CB3 0HA, UK
\and Clarendon Laboratory, University of Oxford, Oxford OX1 3PU, UK}
\titlerunning{`Orthogonal' polarization modes}
\maketitle

\begin{abstract}
We evaluate the Li\'enard-Wiechert field of a rotating superluminal point source numerically 
and show that this radiation field has the following intrinsic characteristics. (i) It is 
sharply focused along a narrow, rigidly rotating spiral-shaped beam that embodies the cusp 
of the envelope of the emitted wave fronts.  (ii) It consists of either one or three 
concurrent polarization modes (depending on the relative positions of the observer and 
the cusp) that constitute contributions to the field from differing retarded times.  (iii) Two 
of the modes are comparable in strength at both edges of the signal and dominate over the third 
everywhere except in the middle of the pulse.  (iv) The position angle of the total field swings 
across the beam by as much as 180$^\circ$.  (v)  The position angles of its two dominant modes 
remain approximately orthogonal throughout their excursion across the beam.  Given the fundamental 
nature of the Li\'enard-Wiechert field, the coincidence of these characteristics with those of 
the radio emission that is received from pulsars is striking.  
\end{abstract}

\section{Introduction}

The rigidly rotating distribution of the radiation from pulsars (reflected in the highly stable periodicity of the mean profiles of the observed pulses) can only arise from a source whose distribution {\it pattern} correspondingly rotates rigidly, a source whose average density depends on the azimuthal angle $\varphi$ in the combination $\varphi-\omega t$, where $\omega$ is the angular frequency of rotation of the pulsar and $t$ is time.  Maxwell's equations demand that the charge and current densities that give rise to this radiation should have exactly the same symmetry ($\partial/\partial t=-\omega\partial/\partial\varphi$) as that of the observed radiation fields ${\bf E}$ and ${\bf B}$.  The domain of applicability of such a symmetry cannot be localized; its boundaries expand.  Unless there is no plasma outside the light cylinder, therefore, the macroscopic distribution of electric current in the magnetosphere of a pulsar should have a superluminally rotating pattern in $r>c/\omega$ (where $r$ is the radial distance from the axis of rotation).  Moving sources of the electromagnetic radiation whose speeds exceed the speed of light {\it in vacuo} have already been generated in the laboratory (\cite{BGMP2004, AHSAFH2004, BS2005}).
%(Bessarab et al.\ 2004, Ardavan et al.\ 2004a, Bolotovskii \& Serov 2005).  
These sources arise from separation of charges: their superluminally moving distribution patterns are created by the coordinated motion of aggregates of subluminally moving particles.  Nevertheless, the radiation that is emitted by a polarization current of this kind is no different from that which is emitted by a similarly-distributed current of free charges (\cite{G1972,BB1990,BS2005}).
%(Ginzburg 1972, Bolotovskii \& Bykov 1990, Bolotovskii \& Serov 2005).\par

We have shown elsewhere that the radiation field of a superluminally rotating extended source at a given observation point $P$ arises almost exclusively from those of its volume elements that approach $P$, along the radiation direction, with the speed of light and zero acceleration at the retarded time (\cite{AAS2004}).  These elements comprise a filamentary part of the source whose radial and azimuthal widths become narrower (like ${R_P}^{-2}$ and ${R_P}^{-3}$, respectively), the larger is the distance $R_P$ of the observer from the source, and whose length is of the order of the length scale of distribution of the source parallel to the axis of rotation (\cite{AAS2006}).  Because these contributing elements are at the same optical distance from the observer, their field in the far zone is effectively the same as that of a point-like source.

The Li\'enard-Wiechert field of a circularly moving point source with the charge $q$ is given by
\begin{eqnarray}
{\bf E}({\bf x}_P,t_P) & = q\sum_{t_{\rm ret}}\Bigg[&{(1-\vert{\dot{\bf x}}\vert^2/c^2){\bf u}\over\vert1-{\hat{\bf n}}\cdot{\dot{\bf x}}/c\vert^3R^2(t)} \nonumber \\
                       &                            & \mbox{} +{{\hat{\bf n}}{\bf\times}({\bf u}{\bf\times}{\ddot{\bf x}})\over c^2\vert1-{\hat{\bf n}}\cdot{\dot{\bf x}}/c\vert^3R(t)}\Bigg],
\end{eqnarray}
and ${\bf B}={\hat{\bf n}}{\bf\times}{\bf E}$, in which
$${\bf x}={\bf x}(t):\qquad r={\rm const.},\quad \varphi=\omega t,\quad z=0,$$
is the trajectory of the source in terms of the cylindrical polar coordinates $(r,\varphi,z)$ based on the axis of rotation.  Here, $({\bf x}_P,t_P)$ are the space-time coordinates of the observation point, ${\bf R}(t)\equiv{\bf x}_P-{\bf x}$, ${\dot{\bf x}}\equiv d{\bf x}/dt$, ${\bf u}\equiv{\hat{\bf n}}-{\dot{\bf x}}/c$, and the unit vector ${\hat{\bf n}}\equiv{\bf R}/R$ designates the radiation direction.  The summation extends over all values of the retarded time, i.e.\ all solutions $t_{\rm ret}<t_P$ of
\begin{eqnarray}
t_P & = & t+R(t)/c \nonumber \\
    & = & t+[{z_P}^2+{r_P}^2+r^2-2rr_P\cos(\varphi_P-\omega t)]^{1/2}/c.
\end{eqnarray}
The field described by Eqs.\ (1) and (2) is identical to that which is encountered in the analysis of synchrotron radiation, except that its value at any given observation time receives contributions from more than one retarded time (\cite{AAS2004}).

\begin{figure}
\centerline{\psfig{file=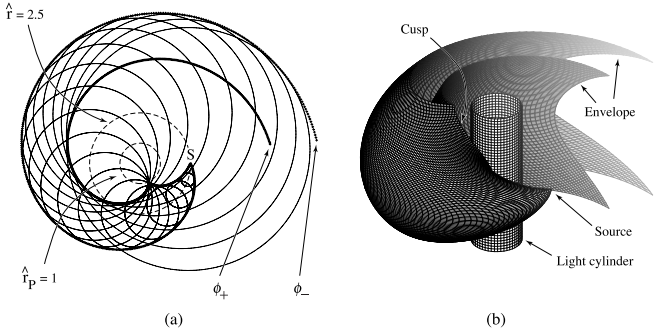,width=8.8cm,clip=} }
\caption{The two-sheeted ($\phi_\pm$) envelope of wave 
 fronts and its cusp; (a) is the cross section of (b) with the plane of 
 source's orbit.  The dimensionless variables ${\hat r}$ and ${\hat r}_P$ 
 are the radial coordinates of the source point and the observation point 
 in units of the light-cyliner radius $c/\omega$. 
 \label{ard1}}
\end{figure}

\begin{figure}
\centerline{\psfig{file=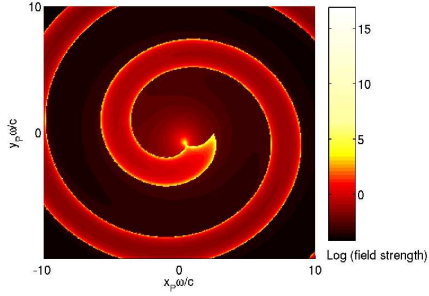,width=6cm,clip=}}
\caption{Field strength in the plane of source's orbit,
for $\hat{r}= 2.5$.  Note the high strength of the field
along the inner edges of the envelope
and near the cusp.
\label{ard3}}
\end{figure}

\begin{figure}
\centerline{\psfig{file=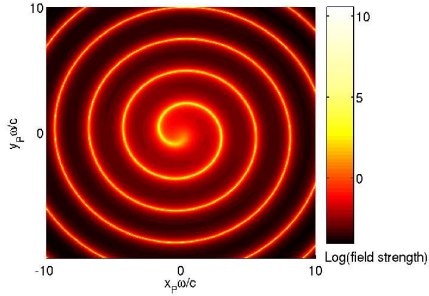,width=6cm,clip=} }
\caption{The distribution of field strength on the cone
$\theta_P = \arcsin(\hat{r}^{-1})$ containing
the cusp.
\label{ard4}}
\end{figure}

In the superluminal regime $r\omega>c$, the wave fronts that are emitted by the above circularly 
moving point source possess a two-sheeted envelope: a tube-like surface whose two sheets meet 
tangentially along a spiraling cusp curve (Fig.~\ref{ard1}).  For moderate superluminal speeds, 
the field inside the envelope receives contributions from three distinct values of the retarded 
time, while the field outside the envelope is influenced by only a single instant of emission time.  
The constructive interference of the emitted waves on the envelope (where two of the contributing 
retarded times coalesce) and on its cusp (where all three of the contributing retarded times coalesce) 
gives rise to the divergence of the Li\'enard-Wiechert field on these loci (\cite{AAS2003}).  
Here we plot the spatial distribution of field (1), excising the narrow regions in which the 
magnitude of this field exceeds a certain threshold (Figs.~\ref{ard3} and \ref{ard4}). 

\begin{figure}
\centerline{\psfig{file=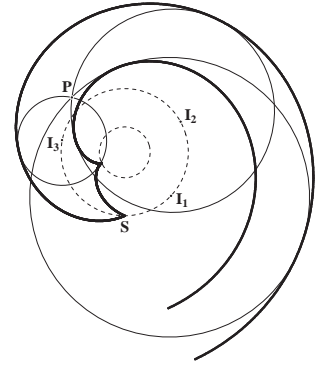,width=4cm,clip=} }
\caption{The three images of the source, emitted at the three 
 contributing retarded times.  An observer $P$ inside the envelope 
 simultaneously receives three wave fronts each of which is emitted 
 at a different retarded time.  The corresponding retarded positions 
 $I_1$, $I_2$ and $I_3$ of the source (the centers of the intersecting 
 wave fronts) are observed as three distinct images of this source.
 \label{ard2}}
\end{figure}

We also consider the individual contributions toward the magnitude and orientation of the total 
field from each of the retarded times.  These contributions, i.e.\ the fields that arise from 
different images of the source (Fig.~\ref{ard2}), each have their own characteristics.  
Though received concurrently, they are emitted at different times and locations and so represent 
distinct radiation modes that can be detected individually (\cite{BB1990}). 

\section{Field strength and polarization}
\begin{figure}
\centerline{\psfig{file=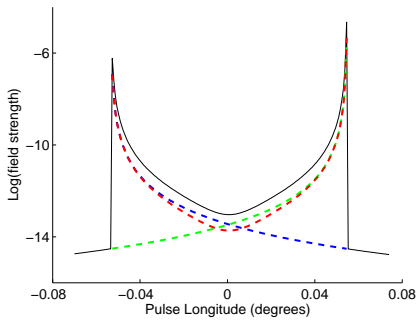,width=6cm,clip=} }
\caption{The relative strengths of the three radiation modes as 
 observed near the cusp on a sphere of large radius. The total 
 field strength (black) and strengths of the underlying contributions 
 from the three images of the source (green, red, blue) are shown 
 for a source with $\hat{r}= 1.1$ and an observation point that 
 sweeps a small arc of the circle $\hat{R}_P\equiv R_P\omega/c=10^{10}$, 
 $\theta_P = \pi/2.7$, crossing the envelope near the cusp. Note that 
 the contribution from the third retarded time (blue) is much stronger 
 than that from the first (green) near the beginning of the pulse, 
 with these r\^{o}les reversed near the end.  Note also that two of the 
 contributions are stronger than the third everywhere except in the middle 
 of the pulse. 
 \label{ard5}}
\end{figure}

\begin{figure}
\centerline{\psfig{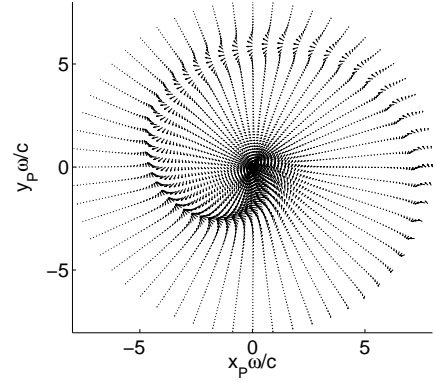}}
\caption{Polarization position angles and field strengths for a source with 
${\hat r}=2$ on the cone $\theta_P = \pi/12$ outside the envelope.
\label{ard6}}
\end{figure}

\begin{figure}
\centerline{\psfig{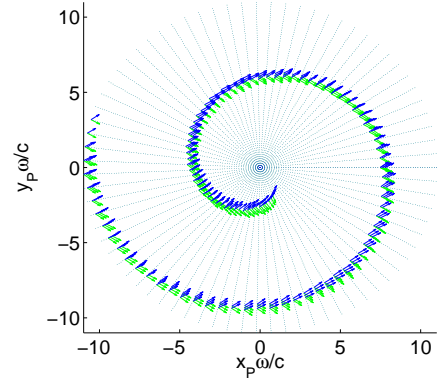} }
\caption{Polarization position angles of the two stronger images.  The arrows represent the directions (but not the magnitudes) of the dominant two of the three simultaneously received contributions to the electric field for ${\hat r}=2$ on the cone $\theta_P = \pi/4$.  Note that the position angles of these two concurrent polarization modes are approximately orthogonal everywhere. 
\label{ard7}}
\end{figure}

\begin{figure}
\centerline{\psfig{file=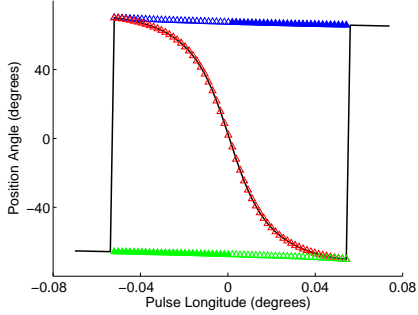,width=6cm,clip=} }
\caption{Position angles of the contributions from the three retarded times (green, red, and blue triangles) are shown relative to one another and to that of the total field (black line), for the same source and observation arc
as in Fig.~\ref{ard5}.
The position angles of the dominant contributions
are shown with open triangles, and those of the weakest contribution with filled triangles.
Note that the position angles of the first (blue) and third (green) contributions differ by approximately $140^\circ$ across the pulse,
and that the position angle of the second contribution (red) closely follows the average position angle,
bridging the first and third contributions.
\label{ard8}}
\end{figure}

\begin{figure}
\centerline{\psfig{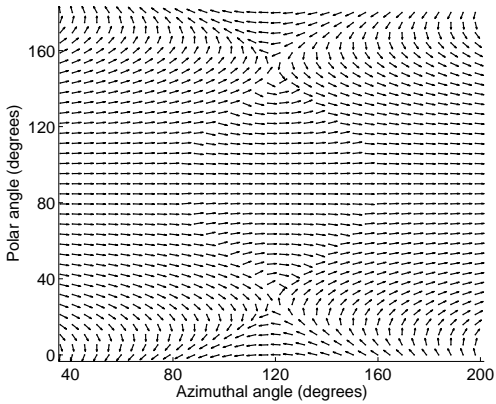} }
\caption{Polarization position angles of the total field on a sphere ${\hat R}_P=10^{10}$, for ${\hat r}=2$.  
The arrows only indicate the direction and not the magnitude of the field.  
The fields sampled by an observer in the course of each rotation are those along a horizontal line.  
Note the $180^\circ$ swings of the position angle in the vicinities of the two points at 
which the cusp intersects the sphere and the field is strongest.
\label{ard9}}
\end{figure}

\begin{figure}
\centerline{\psfig{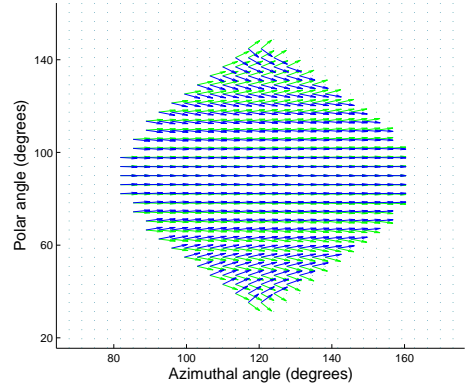} }
\caption{Polarization position angles of the two dominant polarization modes 
(inside the envelope) for the same parameters as those in Fig.\ 9.
\label{ard10}}
\end{figure}

Because the field of the rotating source
itself rotates rigidly,
an observer at spherical coordinates $(R_P,\varphi_P,\theta_P)$
samples, during each rotation period,
the field on the arc that lies at the intersection of the cone $\theta_P =$ constant and the sphere 
$R_P =$ constant.  In Figs.~\ref{ard5}, \ref{ard6}, \ref{ard7}, \ref{ard8}, \ref{ard9} and \ref{ard10}, 
we have illustrated various characteristics of the emission by plotting field strength and 
polarization in the vicinity of the cusp (Fig.~\ref{ard1}). The cusp is tangential to the 
light cylinder at a point that lies in the plane of the source's orbit and spirals upward 
and outward (and symmetrically, downward and outward) from this point, approaching the double 
cone $\theta_P = \arcsin({\hat r}^{-1})$ in the far zone.  It intersects the sphere $R_P =$ 
constant at two points that are symmetrically located with respect to the plane of source's orbit.  

The cross section of a cone whose opening angle $\theta_P$ is slightly larger than 
$\arcsin({\hat r}^{-1})$ with the sphere $R_P =$ constant consists of an arc, close to the 
cusp, that lies partly inside and partly outside the envelope.  Owing to the discontinuous 
change in the strength of the field across the envelope (\cite{AAS2004}), the intensity of 
the radiation sampled along such an arc has a pulsed distribution consisting of two sharply 
peaked components; the closer the opening angle of the cone to $\arcsin({\hat r}^{-1})$, 
the narrower is the width of this two-component pulse (Fig.~\ref{ard5}). 

A quantitative comparison of the features depicted in Figs.\ (3)--(10) (and enumerated in the abstract) with those of the radio emission from pulsars is outside the scope of this report and will be presented elsewhere.  However, the striking similarity of the results shown in these figures with observational data from pulsars cannot be questioned (cf.\ \cite{SCR84}).

\section{Concluding remarks}

The divergence of the magnitude of the Li\'enard-Wiechert field on the envelope of wave fronts and its cusp is a reflection of the fact that no superluminal source can be point-like. When the Li\'enard-Wiechert fields of the constituent volume elements of an extended source are superposed, one obtains a finite field, but a field whose strength diminishes with the distance $R_P$ from the source like ${R_P}^{-1/2}$, rather than ${R_P}^{-1}$, within the bundle of cusps that emanate from these elements (\cite{AAS2004}).  This effect has been observed in the laboratory (\cite{AHSAFH2004}).  The nonspherically decaying radiation arises almost exclusively from those volume elements that approach the observer along the radiation direction with the speed of light and zero acceleration at the retarded time.  

Such elements occupy, at any given observation time, a filamentary part of the source whose transverse dimensions are of the order of $\delta{\hat r}\sim{{\hat R}_P}^{-2}$ in the radial direction and $\delta\varphi\sim{{\hat R}_P}^{-3}$ in the azimuthal direction.  The corresponding dimensions of the bundle of cusps that emanate from the contributing source elements occupy a solid angle in the space of observation points whose azimuthal width $\delta\varphi_P$ has the constant value $\delta\varphi=\delta\varphi_P$ (i.e.\ is subject to diffraction as in a conventional radiation beam) but whose polar width $\delta\theta_P$ decreases with the distance $R_P$ like ${R_P}^{-1}$ (\cite{AAS2006}).  Thus, the area ${R_P}^2\sin\theta_P\delta\theta_P\varphi_P$ subtended by the bundle of cusps defining this subbeam increases like $R_P$, rather than ${R_P}^2$ with the distance $R_P$ from the source, as required by the conservation of energy.

The overall radiation beam within which the nonspherically decaying radiation is detectable has the constant width $\arcsin(1/{\hat r}_>)\le\theta_P\le\arcsin(1/{\hat r}_<)$ for a source with the radial boundaries ${\hat r}_<>1$ and ${\hat r}_>>{\hat r}_<$.  When the length scale of spatial variations of the source is comparable to ${{\hat R}_P}^{-2}$ (e.g.\ in the case of a turbulent plasma with a superluminally rotating macroscopic distribution), the overall beam consists of an incoherent superposition of the coherent nondiffracting subbeams described above (\cite{AAS2006}).

Therefore, in addition to the polarization of this radiation, the nonspherical decay of its intensity (i.e.\ its high brightness temperature), its broad spectrum (\cite{AAS2003}), and the narrowness of both the beam into which it propagates and the region of the source from which it arises, are features of the emission from a rotating superluminal source that are {\it all} fully consistent with observational data from pulsars. 

\begin{acknowledgements}
This work is supported by U.\ S.\ Department of Energy grant LDRD 20050540ER. AA is supported by the Royal Society.
\end{acknowledgements}

%\end{document}
          \clearpage

\end{document}